\documentclass[tightenlines,superscriptaddress,twocolumn,aps,notitlepage,nofootinbib,preprintnumbers]{revtex4-1}

\usepackage{graphicx}
\usepackage{amsmath}
\usepackage{amsfonts}
\usepackage{amssymb}
\usepackage{float}
\usepackage{hyperref}
\usepackage{multirow}
\usepackage{bbold}
\usepackage[usenames]{color}
\usepackage{soul}
\usepackage{makecell}
\usepackage{minibox}
\usepackage{fancyhdr}
\usepackage{enumitem}

\newcommand{\CGMF}{\mathtt{CGMF}}
\newcommand{\Unf}{^{235}\mathrm{U}(n_\mathrm{th},f)}
\newcommand{\Punf}{^{239}\mathrm{Pu}(n_\mathrm{th},f)}

\newcommand{\TKE}{\mathrm{TKE}}
\newcommand{\TXE}{\mathrm{TXE}}
\newcommand{\overbar}[1]{\mkern 1.5mu\overline{\mkern-1.5mu#1\mkern-1.5mu}\mkern 1.5mu}

\begin{document}

\preprint{LA-UR-17-31000}
\title{Hauser-Feshbach fission fragment de-excitation with calculated macroscopic-microscopic mass yields}

\author{Patrick Jaffke}
\email{corresponding author: pjaffke@lanl.gov}
\affiliation{Theoretical Division, Los Alamos National Laboratory, Los Alamos, NM 87545, USA}
\author{Peter M\"{o}ller}
\affiliation{Theoretical Division, Los Alamos National Laboratory, Los Alamos, NM 87545, USA}
\author{Patrick Talou}
\affiliation{Theoretical Division, Los Alamos National Laboratory, Los Alamos, NM 87545, USA}
\author{Arnold J. Sierk}
\affiliation{Theoretical Division, Los Alamos National Laboratory, Los Alamos, NM 87545, USA}

\date{\today}

\begin{abstract}
The Hauser-Feshbach statistical model is applied to the de-excitation of primary fission fragments
using input mass yields calculated with macroscopic-microscopic models of the potential energy surface.
We test the sensitivity of the prompt fission observables to the input mass
yields for two important reactions, $\Unf$ and $\Punf$, for which good experimental data exist. General traits of the
mass yields, such as the location of the peaks and their widths, can impact both the
prompt neutron and $\gamma$-ray multiplicities, as well as their spectra. Specifically, we use
several mass yields to determine a linear correlation between the calculated
prompt neutron multiplicity $\bar{\nu}$ and the average heavy-fragment mass $\langle A_h\rangle$
of the input mass yields $\partial\bar{\nu}/\partial\langle A_h\rangle = \pm 0.1\,n/f/\mathrm{u}$. The mass peak
width influences the correlation between the total kinetic energy of the fission fragments and the total number
of prompt neutrons emitted $\bar{\nu}_T(\TKE)$																																																																																																																																																																																																																																																																																																																																																																																																																									. Typical biases on prompt particle observables from using calculated mass yields instead of
experimental ones are: $\delta \bar{\nu} = 4\%$ for the average prompt neutron multiplicity,
$\delta \overbar{M}_\gamma = 1\%$ for the average prompt $\gamma$-ray multiplicity,
$\delta \bar{\epsilon}_n^\mathrm{LAB} = 1\%$ for the average outgoing neutron energy, $\delta \bar{\epsilon}_\gamma = 1\%$
for the average $\gamma$-ray energy, and $\delta \langle\TKE\rangle = 0.4\%$ for the average total kinetic energy
of the fission fragments.
\end{abstract}

\maketitle


\section{\label{sec:Intro}Introduction}

Nearly 80 years have passed since Hahn and Stra{\ss}mann observed fission products
following the bombardment of uranium with neutrons~\cite{Hahn1939,Meitner1939}. The
data were explained by Meitner and Frisch as the
result of a division of a nucleus into two fragments using an analogy with
a liquid drop~\cite{Meitner1939}. Shortly after, Bohr and Wheeler put this analogy on a quantitative
footing, allowing them to calculate fission-barrier heights fairly well throughout the nuclear
chart~\cite{Bohr1939_2,Bohr1939}. Since 1938, our theoretical description of fission
has continually improved. For example, fission-barrier saddle-point heights are calculated within $\sim 1\,\mathrm{MeV}$
of the empirical values~\cite{Moller2009}
and realistic descriptions of the fragment mass distributions across the $(N,Z)$-plane
are possible~\cite{Randrup2013}.

Fission begins with the formation of
a compound state~\cite{Bohr1936}. The subsequent process leading to the formation of
separate fragments can be described as an evolution in a potential-energy landscape,
where each location corresponds to a specific nuclear shape. A large number of fragment
excitation energies, shapes, and mass splits result, with different formation probabilities. The fragments
de-excite by neutron and $\gamma$-ray emissions. Beta decay and delayed-neutron
emission follow as these unstable nuclei decay towards $\beta$-stability.
Over the years, considerable advancements have been made
in studies of these different processes. For example, some fragment properties have been
reasonably well reproduced using macroscopic-microscopic descriptions of the potential-energy
surface based on Brownian shape-motion dynamics~\cite{Randrup2011,Moller2015} or
Langevin equations~\cite{Aritomo2014,Sierk2017}, or microscopic models based on effective
nucleon-nucleon interactions in terms of energy-density functionals in an adiabatic approximation~\cite{Schunk2014,Regnier2016}
or with full non-adiabatic effects included~\cite{Goddard2015,Bulgac2016}. In addition,
models of the de-excitation via sequential emission of neutrons and $\gamma$ rays~\cite{Talou2013,Vogt2014,Litaize2015}
have been used to describe various prompt neutron and $\gamma$-ray data. Finally, the
delayed-neutron emission and half-lives via $\beta$-n decays have also been investigated based
on a QRPA treatment of transitions in deformed nuclei~\cite{Mumpower2016}. Despite eight decades of progress in modeling
some of the individual steps from scission to the formation of $\beta$-stable fragments, no complete, cohesive
model tying together the various correlated quantities exists.

In this work, we combine mass yields determined from macroscopic-microscopic descriptions
of the potential-energy surface for the compound nucleus shape and dynamics based on either
the Brownian shape-motion~\cite{Randrup2011} or Langevin approach~\cite{Sierk2017} with
a de-excitation model based on a Monte Carlo implementation of the Hauser-Feshbach
statistical-decay theory~\cite{Becker2013}. Using theoretical models
for the fission-fragment yields is attractive for many reasons.
Most notably, the best-studied fission reactions are restricted to a handful of actinides
at a few incident neutron energies, but recent experimental methods have been used
to probe fragment yields beyond this region~\cite{Schmidt2000,Nishio2017}. Even so,
astrophysical $r$-process calculations would require yields for thousands
of nuclei~\cite{Cote2017}. Additionally, many
yields measurement techniques rely on assumptions about the prompt neutron emission from the
primary fragments~\cite{Schillebeeckx1992,Nishio1998}. Another issue is that the inherent
mass resolution in experimental measurements will smear the true yields and only a few
detector setups have been able to achieve the difficult goal of a resolving power less than
one nucleon~\cite{Schillebeeckx1994,Martin2015,Gupta2017,Tovesson2017}.
By connecting theoretical calculations of the fragment yields with
a de-excitation model, one can both estimate fission observables for unknown reactions
and improve our understanding of current experimental data. We use this connection to
determine correlations between the characteristics of the mass yields and the
prompt neutron and $\gamma$-ray emissions. In this way, experimental measurements
of prompt fission observables can inform the development of more accurate
fission models and de-excitation methods. We utilize two commonly studied fission
reactions, $\Unf$ and $\Punf$, as large amounts of experimental data are available on both the
fragment yields and many prompt fission observables. 

Section~\ref{sec:Intro} introduces the main theoretical components of the
macroscopic-microscopic model and the Hauser-Feshbach treatment. We first compare
calculated and experimental mass yields. Then, we compute in Sec.~\ref{sec:Calc} the prompt neutron and $\gamma$-ray emissions with
both sets of input yields. Comparisons between the prompt
observables, such as the neutron and $\gamma$-ray multiplicities and spectra, are made
and we identify the causes of the observed differences. In Sec.~\ref{sec:Conc}, we conclude
by providing estimates of the biases introduced by using calculated yields instead of
experimental ones and identify future improvements and uses for these theoretical models.

\section{\label{sec:Intro}Theoretical Models}
\subsection{\label{ssec:macromicro}Yield Calculation}
The complete specifications of the yield models used here are in Ref.~\cite{Moller2015,Sierk2017}. 
The Brownian shape-motion model used here represents a generalization of the model
introduced in Ref.~\cite{Randrup2011}. In its original formulation, fission-fragment
yields were obtained as a function of nucleon number $A$. Since it was assumed that
the fragment charge-asymmetry ratios $Z/N$ were both equal to the charge asymmetry of the
compound fissioning system, one also obtained charge yields. After further development the
model now provides the two-dimensional yield $Y(Z,N)$ versus fragment proton and neutron
numbers and takes into account pairing effects in the nascent fragments. To illustrate the
main features of the model we briefly outline the original implementation of Ref.~\cite{Randrup2011}.
The first step is to calculate the nuclear potential energy as a function of a discrete
set of five shape variables, namely elongation, neck diameter, the (different) spheroidal
deformations of the two nascent fragments, and the mass asymmetry of the nascent
fragments. To represent with sufficient accuracy this five-dimensional potential-energy
function based on a discrete set of shapes we calculate the potential energy for more
than five million different shapes. The yield is obtained by calculating random walks in this
potential-energy landscape. A starting point, normally the second minimum, is selected.
At any time during the walk a neighbor point to the current point is randomly selected as
a candidate for the next point on the random trajectory. This becomes the next point on
the trajectory if it is lower in energy than the current point; if it is higher in energy
it may become the next point on the trajectory with probability $\exp(-\Delta V/T)$ where $\Delta V$
is the energy difference between the candidate point and current point. The process is
repeated until a shape with neck radius smaller than a selected value for the scission
radius is reached. The values of the shape parameters and potential energy at this
endpoint are tabulated and a new random walk is started. In this way ensembles of various
scission parameters are obtained. The method and its current extensions are discussed and
benchmarked in Refs.~\cite{Moller2015,Randrup2011,Randrup2011_2,Randrup2013,Moller2017}.
The Langevin model starts from essentially the same macroscopic-microscopic
potential-energy model as the Brownian shape-motion model, while including full dynamical
inertial and dissipative effects on fission trajectories. It is currently limited to the
assumption of a fixed $Z$/$N$ ratio, as were the first implementations of the Brownian
shape-motion model~\cite{Randrup2011}.

\begin{figure}[h]
\centering
\includegraphics[width=\columnwidth]{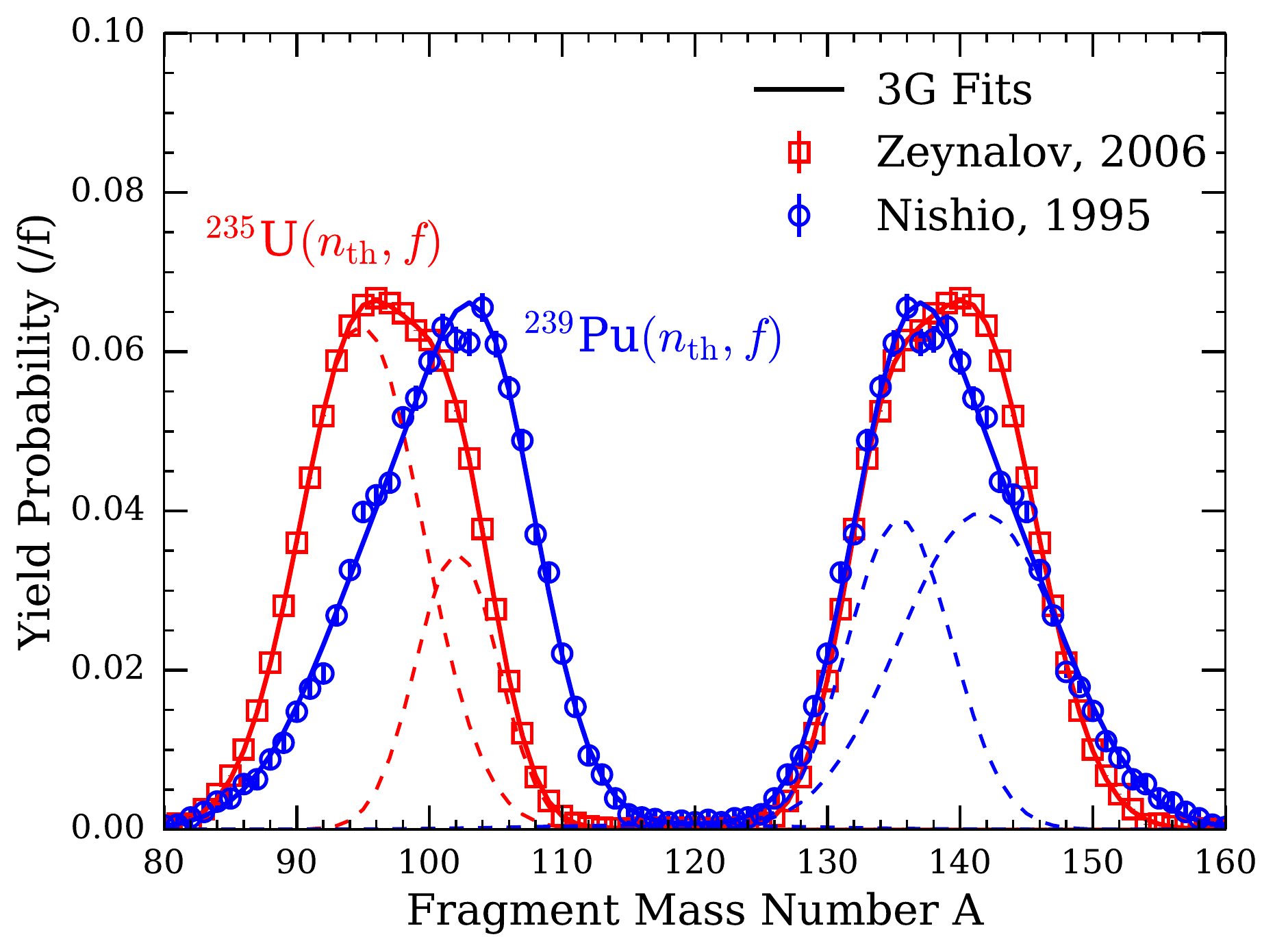}
\caption{\label{fig:3GP} The three-Gaussian parameterization (Eq.~\ref{eq:3GP}) is used to
  fit two experimental mass yields. The squares are for $\Unf$ from Ref.~\cite{Zeynalov2006}
  and the circles are for $\Punf$ from Ref.~\cite{Nishio1995}. The dashed lines are the
  individual Gaussians and the solid curves are the full three-Gaussian fit.
}
\end{figure}

We parameterize the mass yields with the common three-Gaussian
parameterization, similar to the Brosa modes~\cite{Brosa1990}, to generate the
input for a Hauser-Feshbach calculation. The
Gaussians are given by their mean $\mu_i$, variance $\sigma_i^2$, and amplitude $w_i$ as
\begin{equation}
\begin{split}
G_i(A) = \frac{w_i}{\sqrt{2\pi\sigma^2_i}} & \times \bigg[ \exp \bigg( \frac{-(A-\mu_i)^2}{2\sigma^2_i} \bigg) \\
& + \exp \bigg( \frac{-(A-(A_0-\mu_i))^2}{2\sigma^2_i} \bigg) \bigg],
\end{split}
\label{eq:3GP}
\end{equation}
where the indices $i=1,2,3$ refer to the three Gaussians. The Gaussian centered around
the symmetric masses $i=3$ has a fixed mean at $\mu_3 = A_0/2$, with $A_0$ being the
mass of the fissioning nucleus. In addition, the total yields are required to sum to $2$:
$w_1 + w_2 + w_3 = 2$. These requirements reduce the number of variables to seven for
each $Y(A)$. As seen in Fig.~\ref{fig:3GP}, the three-Gaussian fit is an excellent match to
the experimental data
for both the $\Unf$ reaction~\cite{Zeynalov2006} and the $\Punf$ reaction~\cite{Nishio1995}.
We note that the three-Gaussian parameterization is a smooth fit, so it cannot include
shell effects. Nevertheless, this parameterization captures the major aspects of the
mass yields, so we use it as the input mass yields in the de-excitation calculations.

\subsection{\label{ssec:deexcitation}Fragment De-excitation}

The fragment de-excitation process is calculated in the statistical decay theory
of Hauser and Feshbach~\cite{Hauser1952}. In this formalism, the probabilities
for neutron and $\gamma$-ray emission from the excited fragments are calculated at each
stage of the decay. These
probabilities are derived from the transmission coefficients and level densities via
\begin{equation}
\begin{split}
&P(\epsilon_n) d\epsilon_n \propto T_n(\epsilon_n)\rho(A-1,Z,E - \epsilon_n - S_n) d\epsilon_n \\
&P(\epsilon_\gamma) d\epsilon_\gamma \propto T_\gamma(\epsilon_\gamma)\rho(A,Z,E - \epsilon_\gamma) d\epsilon_\gamma,
\end{split}
\label{eq:HFprob}
\end{equation}
where the neutron transmission coefficients $T_n$ are computed using an optical model
with the global optical potential of Koning and Delaroche~\cite{Koning2003}.
The $\gamma$-ray transmission coefficients $T_\gamma$
come from the strength functions in the Kopecky-Uhl formalism~\cite{Kopecky1990} for the
different multipolarities considered. The values for the strength-function parameters
are taken from
the Reference Input Parameter Library (RIPL-3)~\cite{Capote2009}. The level densities
$\rho$ are functions of the fragment mass $A$, charge $Z$, and excitation energy $E$
of the final nuclear state.
They are calculated in the Gilbert-Cameron formalism~\cite{Gilbert1965}, where the low
excitation energy discrete states are used to create a constant temperature model that
connects smoothly to the higher excitation energy continuum states in a Fermi-gas model.
Here, $S_n$ is the neutron separation energy of a fragment with $Z$ protons and $A$ nucleons. Thus,
with Eq.~\ref{eq:HFprob}, one can determine the probability for a given fragment with
excitation energy $E$ to emit either a neutron with energy $\epsilon_n$ or a $\gamma$ ray
with energy $\epsilon_\gamma$. In the Monte Carlo implementation of the Hauser-Feshbach
statistical theory~\cite{Becker2013}, the probabilities are sampled at each step of the
de-excitation until the fragments reach a long-lived isomer or their ground state. This is
done for many fission events resulting in a large data set, where the energy,
spin, and parity are conserved on an event-by-event basis.

To initiate the Hauser-Feshbach decay simulation, one must identify the initial pre-neutron emission
fragment distribution and the excitation energy, spin, and parity distributions. The mass $A$, charge
$Z$, and total kinetic energy $\TKE$ distribution $Y(A,Z,\TKE)$ is sampled to
acquire the initial fragment characteristics of a particular fission event.
The total excitation energy $\TXE$ between the two complementary fragments is then
\begin{equation}
\begin{split}
\TXE = [E_n &+ B_n + M(A_0,Z_0) \\ &- M(A_l,Z_l) - M(A_h,Z_h)] - \TKE(A_h)
\end{split}
\label{eq:Econserv}
\end{equation}
where $l$ and $h$ denote the light and heavy fragment, respectively. The mass and charge
of the fissioning nucleus is $A_0$ and $Z_0$ and, in the case of neutron-induced fission,
$E_n$ is the incident neutron energy and $B_n$ is the binding energy of the target. Thus,
the first term on the right-hand side in Eq.~\ref{eq:Econserv}, represents the Q-value of
the reaction, with $M(A,Z)$ being the mass of a nucleus with mass number $A$ and charge $Z$.

Next, the $\TXE$ is shared between the two fragments. There are several proposed
methods of doing this~\cite{Litaize2010,Schmidt2010,Schmidt2011} and the choice of method
can dramatically affect some fission observables, particularly the average prompt neutron
multiplicity as a function of the fragment mass $\bar{\nu}(A)$~\cite{Stetcu2014,Tudora2015}.
We use the $\CGMF$ code~\cite{CGMF}, which is described in Ref.~\cite{Kawano2010,Becker2013},
to perform the Monte Carlo treatment of the Hauser-Feshbach decay. The TXE is
shared via a ratio of nuclear temperatures $R_T$ with
\begin{equation}
R_T^2 = \frac{T_l^2}{T_h^2} \approx \frac{E_l a_h}{E_h a_l},
\label{eq:RT}
\end{equation}
where the approximation assumes a Fermi-gas model for the level density to relate
the energy $E_i$ to the level-density parameter $a_i$ and the temperature $T_i$. With
$\TXE = E_l + E_h$ and rearranging Eq.~\ref{eq:RT}, we have
\begin{equation}
E_h = \TXE \frac{a_h}{R_T^2 a_l + a_h}.
\label{eq:XEshare}
\end{equation}
The level density parameters depend on the excitation energy of the corresponding
fragments $a_i \equiv a_i(E_i)$, so we iteratively solve the right-hand side of Eq.~\ref{eq:XEshare}
with a given $E_l$ and $E_h$ and corresponding $a_l$ and $a_h$, then adjust $E_l$ and
$E_h$ until the chosen $R_T$ value is satisfied. In general, $R_T$ can have a mass
dependence: $R_T \equiv R_T(A)$. While adjusting $R_T(A)$ in order to reproduce the
experimental $\bar{\nu}(A)$, we have found that it has little impact on our results.

One key ingredient for the simulation is the initial spin distribution of the fission
fragments. As the Hauser-Feshbach model conserves angular momentum, the spin and parity
are needed in order to match levels through $\gamma$-ray emission of different
multipolarities. Currently, E1, M1, and E2 transitions are considered in $\CGMF$. The
spin $J$ distribution follows a Gaussian form
\begin{equation}
P(J) \propto (2J+1) \exp \bigg[ \frac{-J(J+1)\hbar^2}{2\alpha T\mathcal{I}_0(A,Z)} \bigg],
\label{eq:Jdistr}
\end{equation}
where $T$ is the nuclear temperature determined from the level density parameter $a$ and
the excitation energy $E$. The term $\mathcal{I}_0(A,Z)$ is the moment of inertia for a
rigid rotor of the ground-state shape of a fragment with a particular mass and charge.
The factor $\alpha$ is a spin-scaling factor, which can be used to adjust the
average spin of the fragments~\cite{Kawano2013}. Previous studies have
shown that $\alpha$ has a significant effect on the average prompt $\gamma$-ray multiplicity and
energy spectrum~\cite{Stetcu2014}, as well as the isomer production ratios~\cite{Stetcu2013}. In
short, increasing the $J$ of the fragments means more $\gamma$-ray emission at the
expense of neutron emission. These additional $\gamma$ rays are usually dipole
transitions in the continuum region and low in energy. Thus, an increase in $\alpha$
increases $\overbar{M}_\gamma$ and softens the overall $\gamma$-ray spectrum. The
additional $\gamma$ rays in the continuum lead to a slightly lower prompt neutron
multiplicity as well. For this work, we assume equal probability for positive and
negative parity in the level density representation of the continuum in the fission
fragments, i.e. $P(\pi) = 1/2$.
\begin{figure*}[t]
\centering
\includegraphics[width=0.95\textwidth]{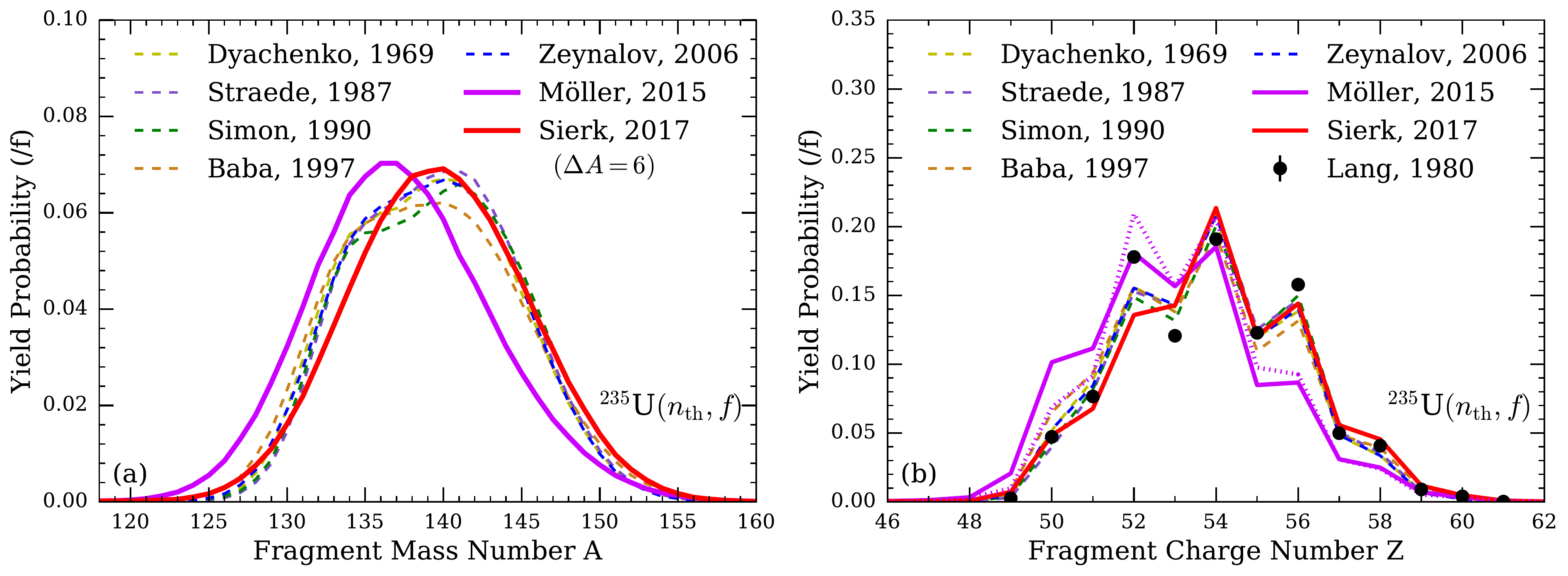}
\caption{\label{fig:YAZU235} (a) The pre-neutron-emission mass yields
  $Y(A)$ for $\Unf$ from the various data sources in the $3$-Gaussian parameterization. The
  colors correspond to the data source. The thick magenta (red) line is the
  result of the theoretical calculation by M\"{o}ller (Sierk) with a $\Delta A = 6.0\,\mathrm{u}$
  mass resolution. (b) The charge yields $Y(Z)$ from the various
  input $Y(A)$ folded with Wahl's~\cite{Wahl2002} $Y(Z|A)$. The dotted line uses the
  $Y(Z|A)$ of M{\"o}ller~\cite{Moller2015}. Black points are the data of Lang {\it et al.}~\cite{Lang1980}.
}
\end{figure*}

\section{\label{sec:Calc}Calculations}

In the past, the de-excitation calculations have sampled from experimental measurements
of the mass yields $Y(A)$, or simple parameterizations~\cite{Vogt2009,Becker2013}. In
this work, we explore the effect on the fission observables from using the calculated
yields described in Sec.~\ref{ssec:macromicro} from Ref.~\cite{Moller2015,Sierk2017}. Our
procedure is straightforward: we conduct the Hauser-Feshbach decay calculations using
$\CGMF$ with different input $Y(A)$ for $\Unf$ and $\Punf$, both of which have a variety
of experimental data available for $Y(A)$ and the various fission observables and
correlations. We perform the calculations with the experimental $Y_e(A)$ and with the
calculated $Y_c(A)$ to determine if there are noticeable effects on the observables. This
sensitivity study is a first step towards determining the predictive capabilities of the calculated
fission yields and developing a fully theoretical and consistent fission model. For
this work, we only study the impact of using the calculated mass yields, and leave the
prospect of using a two-dimensional $Y(A,Z)$ from Ref.~\cite{Moller2015} or a $Y(A,\TKE)$
from Ref.~\cite{Sierk2017} for a future study.

For $\Unf$, we take experimental mass yields $Y_e(A)$ from various data sources~\cite{Dyachenko1969,Straede1987,Simon1990,Baba1997,Zeynalov2006}
and the two calculated mass yields $Y_c(A)$ from M\"{o}ller~\cite{Moller2015} and Sierk~\cite{Sierk2017}.
For $\Punf$, we take $Y_e(A)$ from Ref.~\cite{Wagemans1984,Schillebeeckx1992,Nishio1995,Tsuchiya2000}
and the $Y_c(A)$ from M\"{o}ller~\cite{Moller2015} and Sierk~\cite{Sierk2017}. We use
multiple $Y_e(A)$ in order to determine an uncertainty on the predicted prompt fission observables
simply due to the different input experimental mass yields, which is then compared to
the values obtained with $Y_c(A)$. Input beyond $Y(A)$ are needed to conduct a $\CGMF$
calculation. The calculations require a distribution of fragment charge for a given mass
$Y(Z|A)$, which is taken from the Wahl systematics~\cite{Wahl2002}. One also needs the
average $\TKE$ as a function of the fragment mass $\langle\TKE\rangle (A)$, which we
take from Ref.~\cite{Dyachenko1969} and Ref.~\cite{Wagemans1984} for $\Unf$ and $\Punf$,
respectively. The $R_T(A)$ are deduced in order to best fit $\bar{\nu}(A)$ from Ref.~\cite{Vorobyev2010}
for $\Unf$ and from Ref.~\cite{Apalin1965} for $\Punf$. The $\alpha$ values are
chosen to obtain a reasonable agreement with the $\gamma$-ray multiplicity distributions
of Ref.~\cite{Chyzh2014} and the average $\gamma$-ray multiplicity $\overbar{M}_\gamma$ of
Ref.~\cite{Chyzh2014,Oberstedt2013,Gatera2017}. The total integrated $\TKE$ is allowed to
scale by a factor $\eta$
\begin{equation}
\langle\TKE\rangle = \eta \displaystyle\sum_A \langle\TKE\rangle(A) \times Y(A),
\label{eq:TKE}
\end{equation}
where the sum is over the heavy fragment masses. In our analyses, $\eta$ will be given some
value to scale the calculated $\bar{\nu}$. Typical
values for $\eta$ are within $0.5\%$ of unity. While experimental $\langle\TKE\rangle$ uncertainties are
typically reported as less than $200\,\mathrm{keV}$, these uncertainties are only statistical and the
systematic uncertainties can be closer to $0.6\%$, or $0.5-1.0\,\mathrm{MeV}$~\cite{Milton1962,Wagemans1991,Nishio1995}. Thus, while the
shape of the $\langle\TKE\rangle (A)$ distribution is relatively well-constrained, one
can scale the absolute value more freely. The $\TKE$ for a particular fission event is
sampled from a Gaussian with mean $\langle\TKE\rangle(A)$ and variance $\sigma^2_\TKE(A)$,
which is taken from Ref.~\cite{Zakharova1973} for $\Unf$. For $\Punf$, we use the shape
in Ref.~\cite{Schillebeeckx1992} for $^{240}$Pu(sf). All $\CGMF$ calculations in this study
contain a total of $640000$ fission events.

\begin{figure*}[t]
\centering
\includegraphics[width=0.95\textwidth]{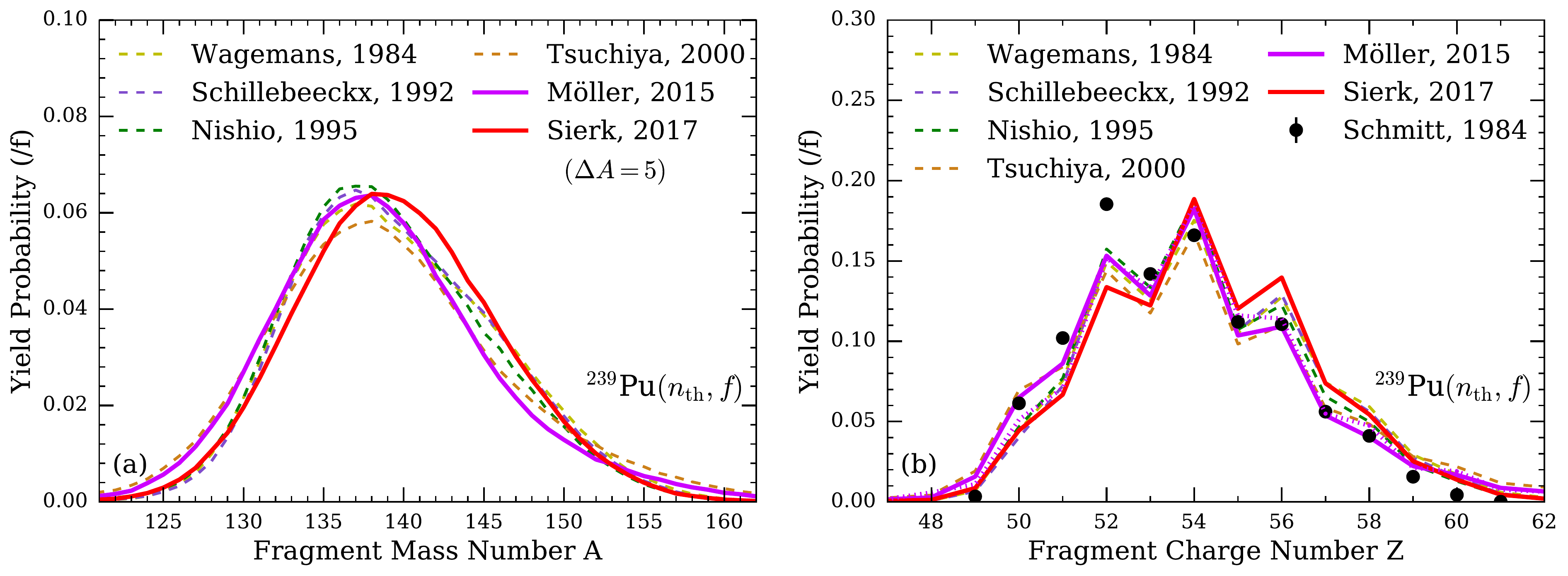}
\caption{\label{fig:YAZPu239} (a) The pre-neutron-emission mass yields
  $Y(A)$ for $\Punf$ from the various data sources in the $3$-Gaussian parameterization. The
  colors correspond to the data source. The thick magenta (red) line is the
  result of the theoretical calculation by M\"{o}ller (Sierk) with a $\Delta A=5.0\,\mathrm{u}$
  mass resolution. (b) The charge yields $Y(Z)$ from the various
  input $Y(A)$ folded with Wahl's~\cite{Wahl2002} $Y(Z|A)$. The dotted line uses
  the $Y(Z|A)$ of M{\"o}ller~\cite{Moller2015}. Black points are the data
  of Schmitt {\it et al.}~\cite{Schmitt1984} using the $Y(A)$ of Ref.~\cite{Schillebeeckx1992}
  as the normalization of the fractional independent yields.
}
\end{figure*}

\begin{table*}[t]
\centering
\renewcommand{\arraystretch}{1.25}
    \begin{tabular}{|c|c|c|c|c|c|c|c|c|c|}
        \hline 
	& Input $Y(A)$ & $\langle A_h\rangle$ (u) & $\sigma^2_{A_h}$ (u$^2$) & $\bar{\nu}$ ($n/f$) & $\langle\nu(\nu-1)\rangle$ & $\langle\nu(\nu-1)(\nu-2)\rangle$ & $\bar{\epsilon}_n^\mathrm{LAB}$ (MeV) & $\overbar{M}_\gamma$ ($\gamma/f$) & $\bar{\epsilon}_\gamma$ (MeV) \\ \hline \hline
	\multirow{7}{*}{$\Unf$} & Dyachenko~\cite{Dyachenko1969}	& 139.16 & 28.69 & 2.458 & 4.766 & 6.749 & 1.984 & 7.284 & 0.856 \\ 
	& Straede~\cite{Straede1987}					& 139.50 & 27.27 & 2.423 & 4.620 & 6.390 & 1.974 & 7.308 & 0.851 \\ 
	& Simon~\cite{Simon1990}					& 139.74 & 30.59 & 2.382 & 4.457 & 6.020 & 1.967 & 7.311 & 0.852 \\ 
	& Baba~\cite{Baba1997}						& 139.00 & 32.88 & 2.458 & 4.772 & 6.783 & 1.988 & 7.274 & 0.860 \\ 
	& Zeynalov~\cite{Zeynalov2006}					& 139.17 & 28.63 & 2.454 & 4.751 & 6.710 & 1.983 & 7.294 & 0.855 \\ \cline{2-10}
	& M\"{o}ller~\cite{Moller2015}					& 137.39 & 33.57 & 2.621 & 5.485 & 8.618 & 2.029 & 7.189 & 0.876 \\ 
	& Sierk~\cite{Sierk2017}					& 139.73 & 31.47 & 2.373 & 4.438 & 6.034 & 1.963 & 7.313 & 0.850 \\ \cline{1-10}
	\multicolumn{2}{|c|}{ENDF/B-VIII.0~\cite{ENDF8}}		&        &       & 2.414$\pm$0.01 & 4.641 & 6.716 & 2.00$\pm$0.01 & 8.19  & 0.89 \\ \hline \hline

	\multirow{6}{*}{$\Punf$} & Wagemans~\cite{Wagemans1984}		& 139.67 & 41.86 & 2.887 & 6.766 & 12.13 & 1.998 & 7.711 & 0.864 \\ 
	& Schillebeeckx~\cite{Schillebeeckx1992}			& 139.61 & 39.65 & 2.901 & 6.828 & 12.31 & 2.001 & 7.715 & 0.863 \\ 
	& Nishio~\cite{Nishio1995}       				& 139.13 & 38.77 & 2.948 & 7.058 & 12.97 & 2.017 & 7.695 & 0.867 \\ 
	& Tsuchiya~\cite{Tsuchiya2000}     				& 139.21 & 57.48 & 2.875 & 6.711 & 12.00 & 2.001 & 7.668 & 0.877 \\ \cline{2-10}
	& M\"{o}ller~\cite{Moller2015}			 		& 138.82 & 48.53 & 2.955 & 7.097 & 13.11 & 2.019 & 7.658 & 0.876 \\ 
	& Sierk~\cite{Sierk2017}					& 139.68 & 38.90 & 2.888 & 6.777 & 12.18 & 1.994 & 7.742 & 0.860 \\ \cline{1-10}
	\multicolumn{2}{|c|}{ENDF/B-VIII.0~\cite{ENDF8}}		&        &       & 2.870$\pm$0.01 & 6.721 & 12.51 & 2.117$\pm$0.037 & 7.33  & 0.87 \\ \hline
     \end{tabular}
\caption{\label{tab:Avgs} Average quantities for $\CGMF$ calculations utilizing different
  mass yields for $\Unf$ and $\Punf$. These calculations used $\langle\TKE\rangle = 171.40\,\mathrm{MeV}$
  for $\Unf$~\cite{Nishio1998} and $\langle\TKE\rangle = 177.93\,\mathrm{MeV}$ for $\Punf$~\cite{Schillebeeckx1992}.
  Listed are the average heavy-fragment mass $\langle A_h\rangle$, the heavy-fragment
  peak variance $\sigma^2_{A_h}$, the average prompt neutron multiplicity $\bar{\nu}$, its first $\langle\nu(\nu-1)\rangle$
  and second $\langle\nu(\nu-1)(\nu-2)\rangle$ factorial moments, the average prompt neutron
  energy in the lab frame $\bar{\epsilon}_n^\mathrm{LAB}$, the average prompt $\gamma$-ray
  multiplicity $\overbar{M}_\gamma$, and average $\gamma$-ray energy $\bar{\epsilon}_\gamma$.
  The calculations used an energy threshold of $\epsilon_n^\mathrm{LAB} > 10\,\mathrm{keV}$ and $\epsilon_\gamma > 100\,\mathrm{keV}$,
  as well as a timing window of $\Delta t = 10\,\mathrm{ns}$ for the $\gamma$ rays. Values
  from ENDF/B-VIII.0~\cite{ENDF8} are also listed with similar detection thresholds.
}
\end{table*}

Pre-neutron-emission fragment mass and charge yields are presented in Fig.~\ref{fig:YAZU235}
for $\Unf$ and in Fig.~\ref{fig:YAZPu239} for $\Punf$. The dashed lines are the three-Gaussian
parameterizations for the different $Y_e(A)$. The thick solid lines are the three-Gaussian
parameterizations for the two $Y_c(A)$. We note that the calculated yields $Y_c(A)$~\cite{Moller2015,Sierk2017}
have been folded with a mass resolution of $\Delta A \sim 6\,\mathrm{u}$ at FWHM.
The resulting charge yields $Y(Z)$ are also given for each reaction.
Recall that the $Y(Z|A)$
are from Wahl~\cite{Wahl2002}, but the differences in the $Y(A)$ are propagated to the
resulting $Y(Z)$, where we see that the spread in the $Y(Z)$ is directly
correlated to that in $Y(A)$. For example, the increase between $125\leq A\leq 135$
for $\Unf$ in the $Y_c(A)$ of Ref.~\cite{Moller2015} is accompanied by an increase in the charge yields
around $49\leq Z\leq 51$. The same trends are found in Fig.~\ref{fig:YAZPu239} for
$\Punf$.

An important feature of the fragment mass yields is the average heavy fragment mass $\langle A_h\rangle$.
From Eq.~\ref{eq:TKE}, one can see that masses with the largest yields, i.e. those near $\langle A_h\rangle$,
will dominate the sum and determine the $\langle\TKE\rangle$ to first order.
The input $\langle\TKE\rangle(A)$ of Ref.~\cite{Dyachenko1969} and Ref.~\cite{Wagemans1984}
both peak near $A=132$. Thus, mass yields with large $Y(A\sim 132)$
will result in the largest $\langle\TKE\rangle$. From Eq.~\ref{eq:Econserv}, we
note that a larger $\langle\TKE\rangle$ results in a lower $\langle\TXE\rangle$, which
provides less energy for the prompt neutron and $\gamma$-ray emissions. In addition, a
different set of mass yields will generate a change in the $Q$-value for the fission
reaction as the fragment masses are different. For our calculations, we have either fixed
the $\langle\TKE\rangle$ to be $171.40\,\mathrm{MeV}$~\cite{Nishio1998} for $\Unf$ and $177.93\,\mathrm{MeV}$~\cite{Schillebeeckx1992}
for $\Punf$, or allowed the $\langle\TKE\rangle$ value to float but restrict $\bar{\nu}$
to be in agreement with the IAEA standards~\cite{Carlson2009}: $2.419\,n/f$ for $\Unf$ and
$2.877\,n/f$ for $\Punf$.

As seen in Table~\ref{tab:Avgs}, the changes in the mass yields can translate to a change
in prompt fission observables. For these $\CGMF$ calculations, we used a fixed $\langle\TKE\rangle$,
which means that the $\eta$ values are different for each choice of $Y(A)$ via Eq.~\ref{eq:TKE}.
This change in $\eta$ shifts the $\langle\TKE\rangle(A)$, which shifts the $\langle\TXE\rangle(A)$
in the opposite direction. Thus, lower $\eta$ values will increase $\langle\TXE\rangle(A)$
and result in a larger $\bar{\nu}$ for the fission reaction, as the excitation energy is
\begin{figure}[t]
\centering
\includegraphics[width=\columnwidth]{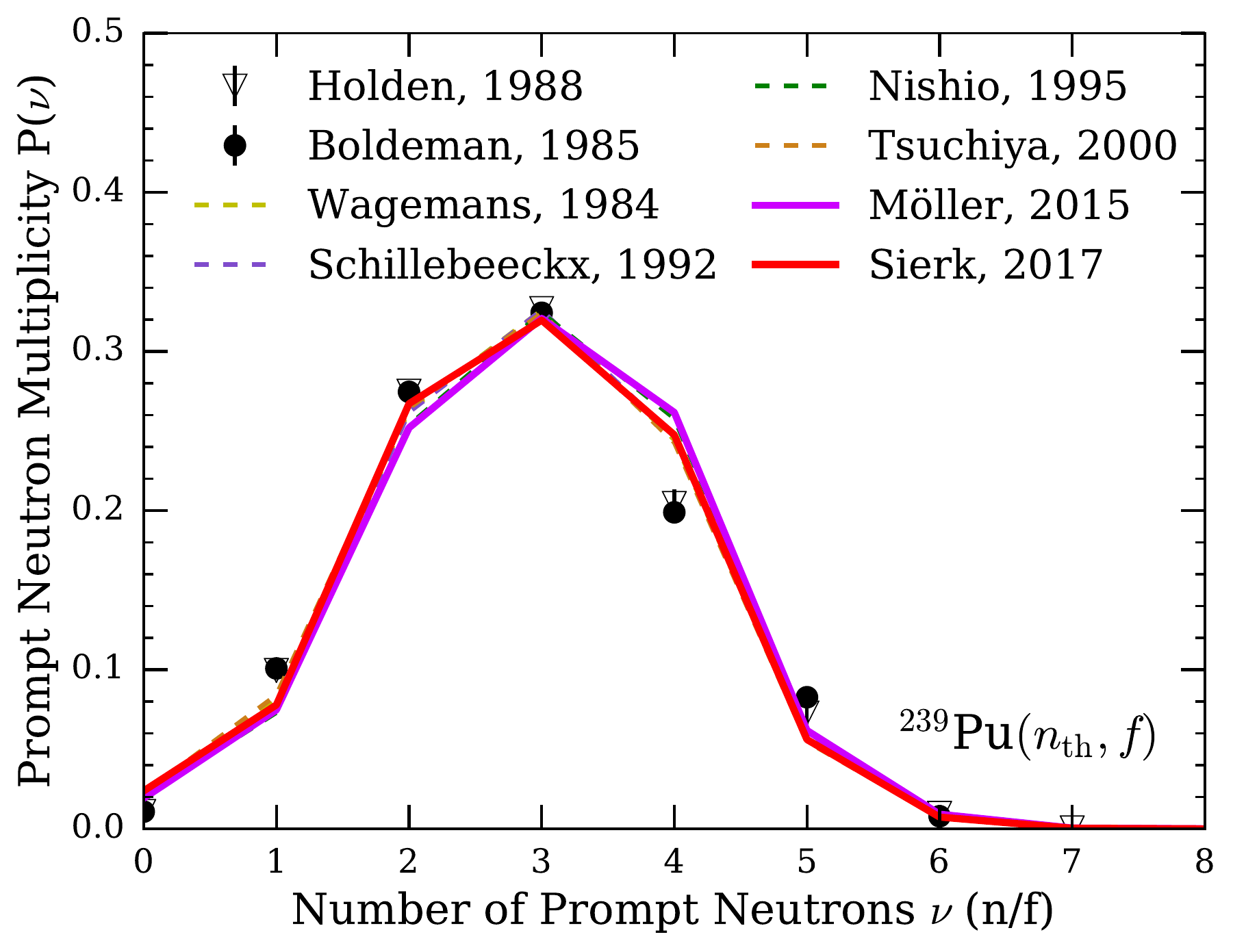}
\caption{\label{fig:Pu239PNu} The prompt neutron multiplicity distribution $P(\nu)$ of
  $\Punf$ for the various input experimental $Y_e(A)$ (dashed lines) or calculated $Y_c(A)$
  (thick solid lines). Black points are the data of Boldeman~\cite{Boldeman1985} and
  Holden~\cite{Holden1988}.
}
\end{figure}
largely removed by neutron emission~\cite{Talou2016}. Assuming $5\,\mathrm{MeV}/n$ from
averaging over all fragments, the statistical differences in the $\langle\TKE\rangle$ values for the
calculations in Table~\ref{tab:Avgs} could only account for a difference of $0.4\%$
in $\bar{\nu}$. However, we find that the different $Y(A)$ can produce up to a $7.7\%$
change in $\bar{\nu}$ and a $1.2\%$ change in $\overbar{M}_\gamma$. This change in $\overbar{M}_\gamma$
is relatively small, compared with the experimental uncertainties~\cite{Chyzh2014,Oberstedt2013,Gatera2017}
and could be solely caused by the correlation between $\bar{\nu}$ and $\overbar{M}_\gamma$~\cite{Wang2016},
i.e. that the change in $\overbar{M}_\gamma$ is only indirectly related to the change in
$Y(A)$ through the change in $\bar{\nu}$. The differences in $\bar{\nu}$, however, are
$1-8\%$, about an order of magnitude larger than the experimental uncertainties~\cite{Boldeman1985,Holden1988}.
This indicates that $\bar{\nu}$ can be very sensitive to the choice of $Y(A)$. The
overall trend in Table~\ref{tab:Avgs} is that a $Y(A)$ with $\langle A_h\rangle$ closer
to $132$ will result in a lower $\eta$ to maintain a fixed $\langle\TKE\rangle$. This will
then increase $\langle\TXE\rangle$ and produce more prompt neutrons. Figure~\ref{fig:Pu239PNu}
demonstrates this point for $\Punf$. We note that the factorial moments of $P(\nu)$ are very sensitive to the variance $\sigma^2_\mathrm{TKE}(A)$.
A scaling of $\sigma^2_\mathrm{TKE}(A)$ by $0.76$
for $\Unf$ and $0.81$ for $\Punf$ was used to obtain reasonable agreement with the
experimental $P(\nu)$~\cite{Boldeman1985,Holden1988}. All calculations shown in this
work use the same $\sigma^2_\mathrm{TKE}(A)$ and the same scaling, meaning that the
changes in $P(\nu)$ seen in Fig.~\ref{fig:Pu239PNu} are a direct result of the change in
$Y(A)$ only.
\begin{figure}[t]
\centering
\includegraphics[width=\columnwidth]{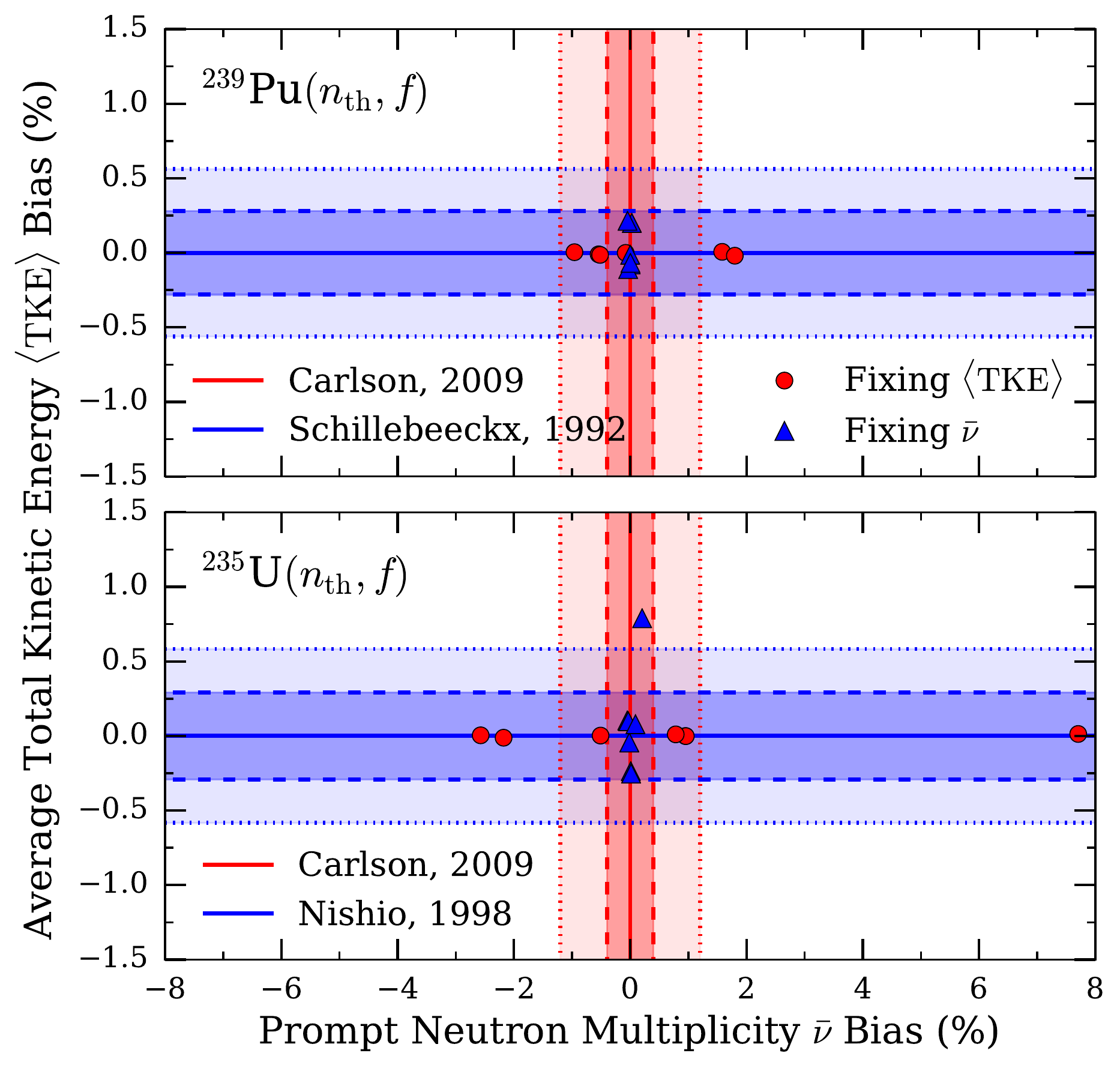}
\caption{\label{fig:ExpVar} The correlation between the average total kinetic
  energy of the fragments $\langle\TKE\rangle$ and the average prompt neutron multiplicity $\bar{\nu}$
  for $\Unf$ (bottom) and $\Punf$ (top). The calculations have either a fixed
  $\langle\TKE\rangle$ (circles) or a fixed $\bar{\nu}$ (triangles).
  Horizontal lines are the experimental $\langle\TKE\rangle$ for
  $\Unf$~\cite{Nishio1998} and $\Punf$~\cite{Schillebeeckx1992} with the shaded regions
  representing $\pm 0.5\,\mathrm{MeV}$ (dashed darker region) and $\pm 1.0\,\mathrm{MeV}$ (dotted lighter region).
  Vertical lines are the evaluated $\bar{\nu}$ and their $1\sigma$
  (dashed darker region) and $3\sigma$ (dotted lighter region) uncertainty bands~\cite{Carlson2009}.
}
\end{figure}

From our initial calculations, we can already see that differences in $Y(A)$ can
produce changes in $\bar{\nu}$ above the sub-percent reported uncertainties for both
$\Unf$ and $\Punf$ evaluated by the standards group~\cite{Carlson2009}. We can invert the
procedure to instead fix $\bar{\nu}$ to the evaluated values and determine the
corresponding $\langle\TKE\rangle$ value needed. This procedure, and its comparison with
the method of fixing $\langle\TKE\rangle$, is shown in Fig.~\ref{fig:ExpVar}.
The different $Y(A)$ induce
typical errors of $\delta\langle\TKE\rangle\sim 0.4\%$ and $\delta\bar{\nu}\sim 4\%$. One
intriguing result from this study is that a highly precise measurement of $\bar{\nu}$
could be used to constrain the allowed values for $\langle\TKE\rangle$, as already mentioned
in Ref.~\cite{Randrup2017}. In the bottom
plot of Fig.~\ref{fig:ExpVar}, when we fix $\bar{\nu}$ to the
evaluated value, the spread in $\langle\TKE\rangle$ values induced from the choice of $Y(A)$
is within the $\pm 0.5\,\mathrm{MeV}$ range. This implies that the experimental
uncertainty on $\langle\TKE\rangle$, $1.4\,\mathrm{MeV}$ in Ref.~\cite{Nishio1998} could
be reduced by the constraints on $\bar{\nu}$ by about a factor of $3$. The differences in
the input mass yields seem to limit this type of correlation analysis to about $\pm 0.4\,\mathrm{MeV}$
in the $\langle\TKE\rangle$ uncertainties. We note that the average spin of the
fragments, governed by $\alpha$, and the shape of $\langle\TKE\rangle(A)$ will also
influence the correlation between $\langle\TKE\rangle$ and $\bar{\nu}$.

\begin{figure}[t]
\centering
\includegraphics[width=\columnwidth]{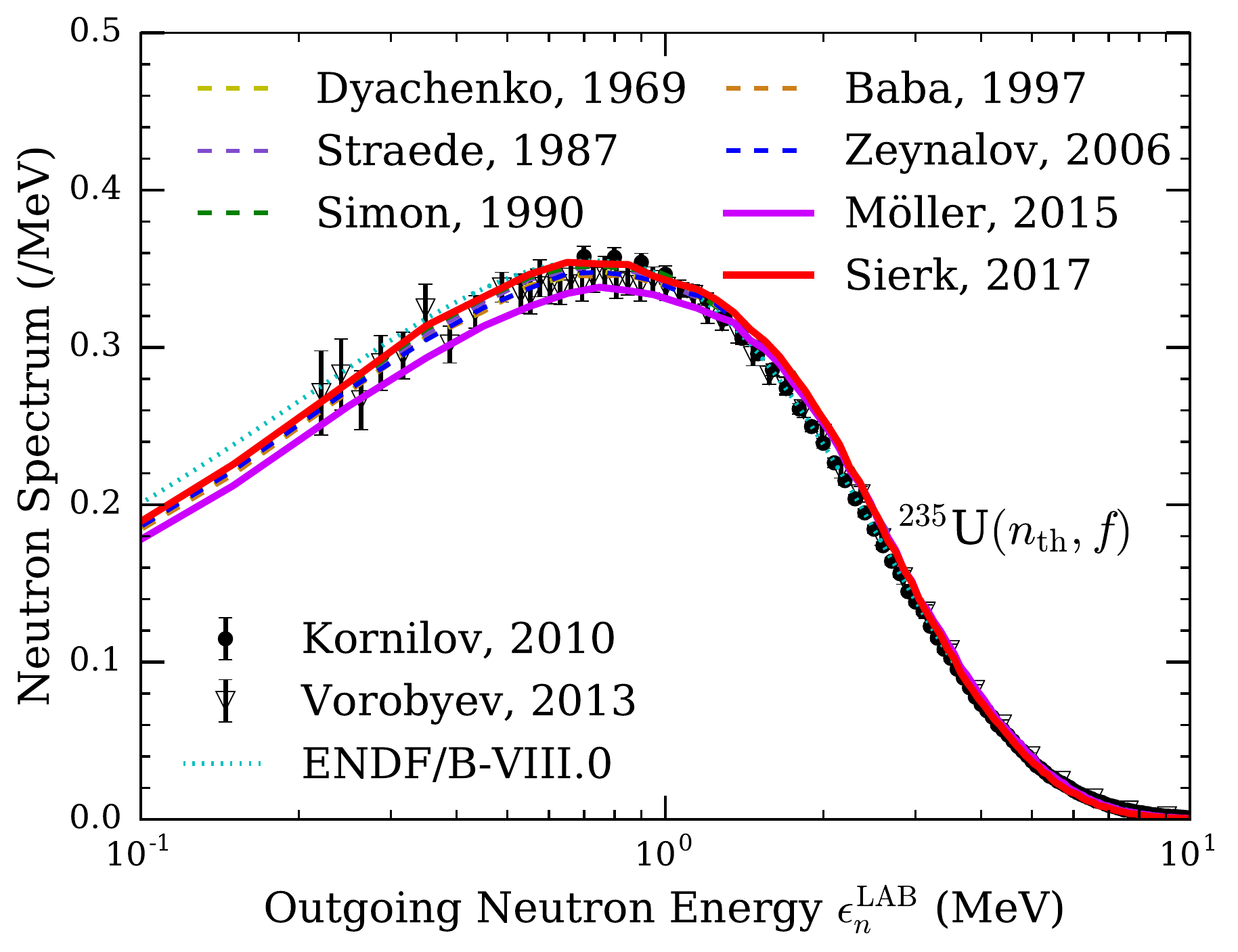}
\caption{\label{fig:U235EnSpec} The prompt fission neutron energy spectrum (PFNS) of $\Unf$ in the lab frame
  calculated with the various input experimental $Y_e(A)$ (dashed lines) or calculated $Y_c(A)$
  (thick solid lines). Black points are the data of Kornilov {\it et al.}~\cite{Kornilov2010}
  and Vorobyev {\it et al.}~\cite{Capote2016}. The dotted line is from ENDF/B-VIII.0~\cite{ENDF8}.
 }
\end{figure}

\begin{figure}[t]
\centering
\includegraphics[width=\columnwidth]{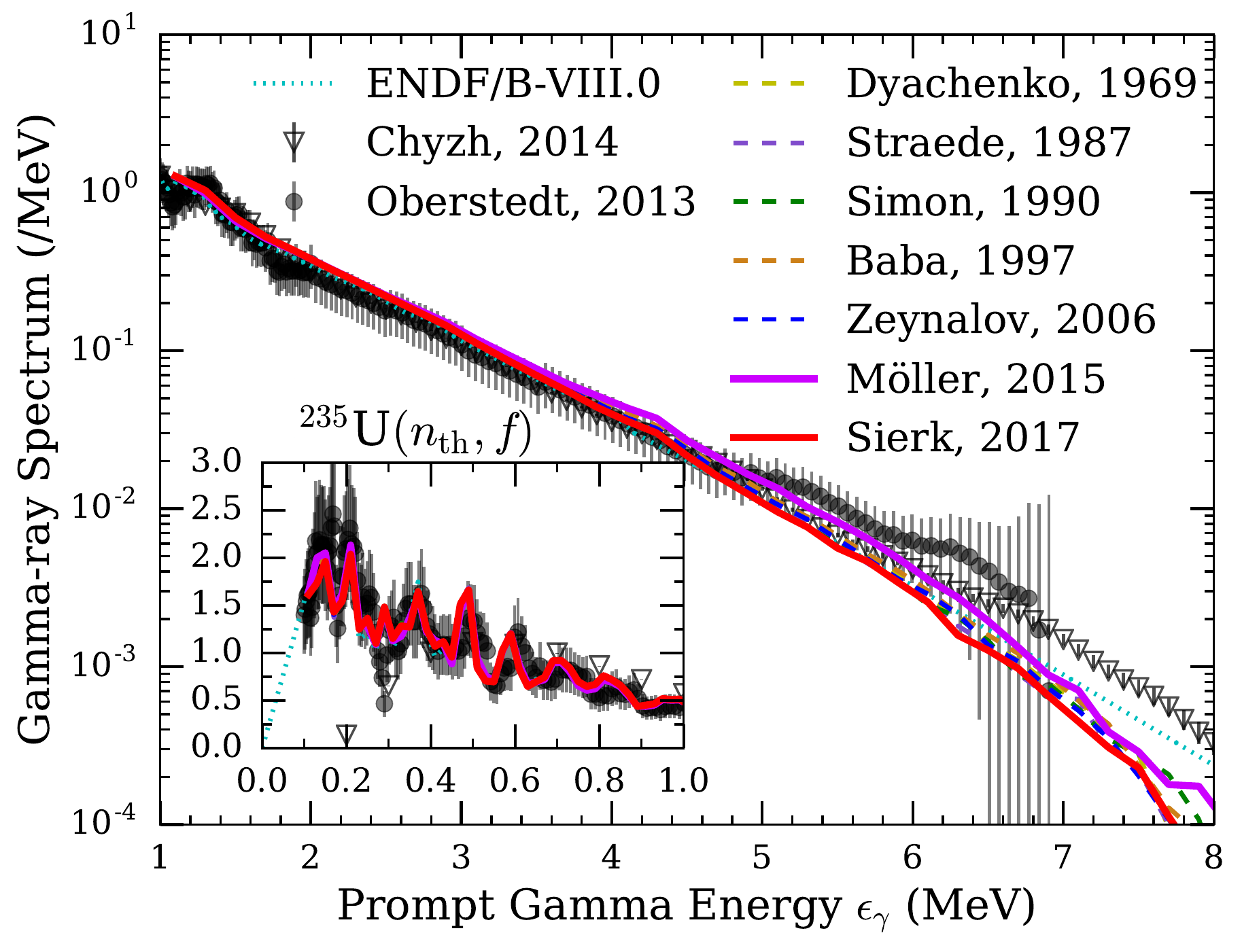}
\caption{\label{fig:U235EgSpec} The prompt fission $\gamma$-ray energy spectrum (PFGS) of
  $\Unf$ calculated with the various input experimental $Y_e(A)$ (dashed lines) or
  calculated $Y_c(A)$ (thick solid lines). Black points are the data of Oberstedt {\it et al.}~\cite{Oberstedt2013}
  and Chyzh {\it et al.}~\cite{Chyzh2014}. We use an energy threshold of $100\,\mathrm{keV}$
  and a timing window of $10\,\mathrm{ns}$ in accordance with Ref.~\cite{Oberstedt2013}.
  The dotted line is from ENDF/B-VIII.0~\cite{ENDF8}. Insert shows the low-energy
  region.
 }
\end{figure}

The changes in the prompt fission neutron (PFNS) and prompt fission $\gamma$-ray (PFGS) spectra are
shown in Fig.~\ref{fig:U235EnSpec} and Fig.~\ref{fig:U235EgSpec}.
The PFNS is plotted to
illustrate the impact of the different $Y(A)$ at low outgoing neutron energies. We can
see that mass yields shifted closer to symmetry will have a slightly harder PFNS, as the average
neutron energies are larger for these masses~\cite{Gook2014,Nishio1998,Tsuchiya2000}. Even with this shift, the
typical error on the average outgoing neutron energy from using calculated mass yields
is $\delta\bar{\epsilon}_n^\mathrm{LAB} \sim 1\%$. Overall, the PFNS is mostly insensitive to the
choice of input $Y(A)$. An additional note is that the PFNS calculated by $\CGMF$
are consistently softer than the experimental ones for neutron
energies above $4\,\mathrm{MeV}$, an issue also identified in previous studies~\cite{Becker2013,Kawano2013}.
This work demonstrates that the choice of input mass yields does not seem to account for this
discrepancy.

The PFGS in Fig.~\ref{fig:U235EgSpec} also appears relatively insensitive to the choice of
input $Y(A)$. We note that the calculation of $\Unf$ using the $Y(A)$ from M{\"o}ller~\cite{Moller2015}
produces a slightly harder PFGS as the mass yields are more shifted towards the $N=82$
closed shell, where the average $\gamma$-ray energy is known to peak~\cite{Pleasonton1972,Hotzel1987}
due to the large level spacing. A similar argument reveals why the average $\gamma$-ray
energy for the Tsuchiya {\it et al.}~\cite{Tsuchiya2000} mass yields is relatively large. Even though
its average heavy fragment peak is not the closest to $A=132$, the peak
width is large enough to produce larger yields for $A\sim 132$ than the other input yields,
as seen in Fig.~\ref{fig:YAZPu239}. Thus, {\it both} $\langle A_h\rangle$ {\it and} $\sigma^2_{A_h}$
can impact the prompt fission observables. We note that specific $\gamma$-ray lines are
sensitive to the choice of input mass yields, as seen in the insert in Fig.~\ref{fig:U235EgSpec}.
For example, the $212.53\,\mathrm{keV}$
peak of $^{100}$Zr is $5\%$ more intense with the $Y_c(A)$ of M{\"o}ller~\cite{Moller2015}
instead of Sierk~\cite{Sierk2017}, due to the change in peak location seen in Fig.~\ref{fig:YAZU235}.
Overall, typical errors of $\delta\bar{\epsilon}_\gamma \sim 1\%$ occur when using the calculated
yields over experimental ones. We note that recent studies involving significantly different
mass yields, such as those between spontaneous fission and neutron-induced fission
from the same compound nucleus, can generate a measurable difference in the
PFGS~\cite{Chyzh2017}. In Fig.~\ref{fig:U235EgSpec},
the calculated spectra deviate from the experimental data above $\epsilon_\gamma = 5\,\mathrm{MeV}$,
with the $\CGMF$ calculations underpredicting the measured spectrum. Previous studies~\cite{Stetcu2013,Stetcu2014}
have demonstrated that decreasing the spin-scaling factor $\alpha$ can increase
the slope of the PFGS, but this will lower $\overbar{M}_\gamma$, creating tension
with the values of Ref.~\cite{Oberstedt2013,Chyzh2014,Gatera2017}.

\begin{figure}[h]
\centering
\includegraphics[width=\columnwidth]{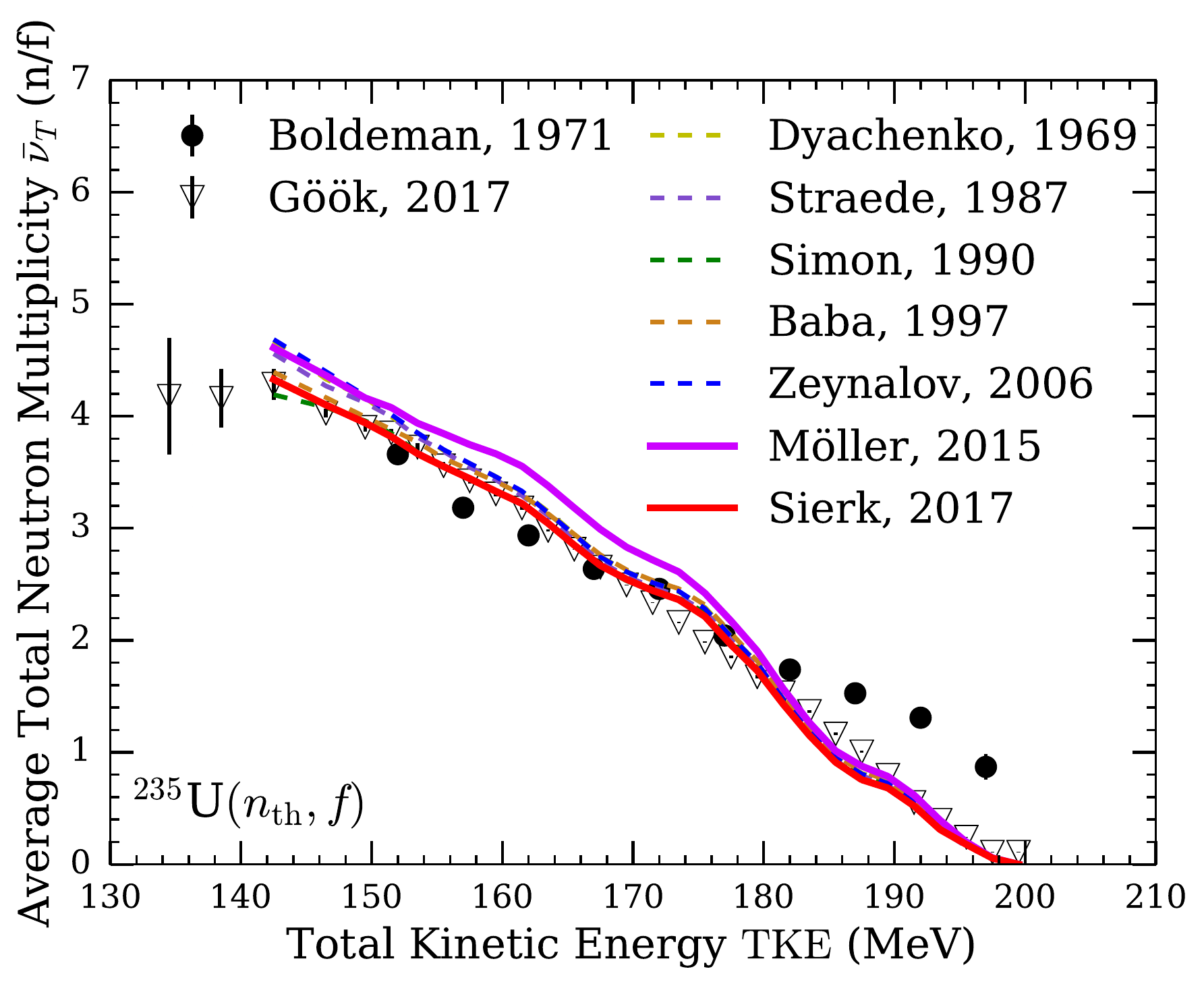}
\caption{\label{fig:U235nubarTTKE} The correlation between the total kinetic
  energy of the fission fragments and the average total prompt neutron
  multiplicity between both the light and heavy fragments $\bar{\nu}_T(\TKE)$ of $\Unf$ for various input
  experimental $Y_e(A)$ (dashed lines) or calculated $Y_c(A)$ (thick solid lines).
  Black points are the data from Boldeman {\it et al.}~\cite{Boldeman1971} and
  G\"{o}\"{o}k {\it et al.}~\cite{Gook2017}.
 }
\end{figure}

We now turn to the correlation between the total kinetic energy and the total number of
prompt neutrons emitted from both the light and heavy fragment $\bar{\nu}_T$. This
relation utilizes the energy conservation in Eq.~\ref{eq:Econserv} and is expected to be
anti-correlated as a larger $\TKE$ results in less energy available for
prompt neutron emission. In Fig.~\ref{fig:U235nubarTTKE}, this trend is seen by the
decreasing trend of $\bar{\nu}_T(\TKE)$ for $\Unf$. The $\CGMF$ calculations
reproduce the experimental data of G\"{o}\"{o}k {\it et al.}~\cite{Gook2017} very well.
A possible explanation for the differences seen for $\TKE>180\,\mathrm{MeV}$
is a broader $\TKE$ resolution in Boldeman {\it et al.}~\cite{Boldeman1971}. The
$\TKE$ bins below $140\,\mathrm{MeV}$ have poor statistics in the $\CGMF$
calculations, so we have cut the calculated curves at this value. We note two trends seen
in Fig.~\ref{fig:U235nubarTTKE}. First, mass yields with a lower $\langle A_h\rangle$ require
a lower $\eta$ to keep $\langle\TKE\rangle$ fixed, which results in more excitation energy
available for the fragments and a shift towards higher $\bar{\nu}_T(\TKE)$.
Second, mass yields with wider peaks (larger $\sigma^2_{A_h}$) result
in a shallower slope for the $\TKE < 160\,\mathrm{MeV}$ bins. For example,
the result using $Y_e(A)$ from Baba {\it et al.}~\cite{Baba1997} is very similar to the result
when using $Y_e(A)$ from Ref.~\cite{Dyachenko1969,Straede1987,Zeynalov2006} for $\TKE > 160\,\mathrm{MeV}$,
but becomes closer to the result using $Y_c(A)$ from Sierk~\cite{Sierk2017} for $\TKE < 160\,\mathrm{MeV}$.
When we take a single $Y(A)$ and arbitrarily add a
mass resolution, which keeps $\langle A_h\rangle$ about constant while increasing $\sigma^2_{A_h}$,
we find the same trend.
This occurs because a larger $\sigma^2_{A_h}$ introduces a wider variety of mass
yields contributing to the same $\TKE$ energy bin. In particular, for the
lower $\TKE$ energy bins, the contribution of very asymmetric yields
increases, which also tend to have a low $\bar{\nu}_T$~\cite{Vorobyev2010}. This
additional influence of very asymmetric mass splits lowers the $\bar{\nu}_T$
for that $\TKE$ energy bin, thus resulting in the trend seen in Fig.~\ref{fig:U235nubarTTKE}.
This low-$\TKE$ region is difficult for experiments, where correcting for detector
effects, such as neutron scattering, capture efficiency, and the recoil imparted onto the
fragment, can play a large role~\cite{Gavron1974,Nifenecker1974,Gook2014,Vorobyev2010}.
Overall, we find that the shift towards higher $\bar{\nu}_T(\TKE)$ is primarily due to the
different $\langle A_h\rangle$, while the change in the slope of $\bar{\nu}_T(\TKE)$ at
low $\TKE$ values is due to the different $\sigma^2_{A_h}$.

\section{\label{sec:Conc}Conclusion}

We have used theoretical models for the fragment mass yields~\cite{Moller2015,Sierk2017}
as input for Hauser-Feshbach simulations of the emission of prompt neutrons and $\gamma$ rays~\cite{Becker2013}.
This allows
us to test the feasibility of using theoretically calculated fission-fragment yields and determine the
sensitivity of important prompt fission observables, such as the average prompt neutron multiplicity $\bar{\nu}$,
average total kinetic energy of the fragments $\langle\TKE\rangle$, and average energies of the
emitted neutrons and $\gamma$ rays, to the input yields. We utilize the $\Unf$ and $\Punf$ reactions, as there
is significant experimental data for both the mass yields $Y(A)$ and prompt fission
observables. An initial comparison of the mass yields demonstrates that the calculated
yields can achieve reasonable agreement with most experimental data. Using a Monte Carlo implementation
of the Hauser-Feshbach statistical decay model~\cite{CGMF}, we propagate the differences
between the experimental and calculated mass yields to differences in the prompt neutron and
$\gamma$-ray observables. In particular, we
find that the average heavy fragment mass $\langle A_h\rangle$ is very influential in
determining $\langle\TKE\rangle$,
which, in turn, is a major factor in determining $\bar{\nu}$.
This finding is reflected in Table~\ref{tab:errors}, where we list the correlation between the
calculated $\bar{\nu}$ and input $\langle A_h\rangle$. The correlation is determined by fitting
ordered pairs of $(\langle A_h\rangle,\bar{\nu})$ for each set of mass yields
in Tab.~\ref{tab:Avgs}.
\begin{table*}
\centering
\renewcommand{\arraystretch}{1.25}
    \begin{tabular}{|c||c|c|c|c|c|c|}
        \hline 
	& \multicolumn{3}{c|}{$\Unf$} & \multicolumn{3}{c|}{$\Punf$} \\ \hline
	$\partial\bar{\nu}/\partial\langle A_h\rangle$ ($n/f$/u) & \multicolumn{3}{c|}{$\pm 0.11$} & \multicolumn{3}{c|}{$\pm 0.08$} \\ \hline \hline
	& M{\"o}ller~\cite{Moller2015} & Sierk~\cite{Sierk2017} & Exp. or Eval. & M{\"o}ller~\cite{Moller2015} & Sierk~\cite{Sierk2017} & Exp. or Eval. \\ \hline \hline
	$\delta\langle\TKE\rangle$ (MeV) 		& $0.8\%$ & $0.3\%$ & $0.6\%$~\cite{Wagemans1991} 	& $0.2\%$ & $0.1\%$ & $0.6\%$~\cite{Wagemans1991} \\ \hline
	$\delta\bar{\nu}$ ($n/f$) 			& $7.7\%$ & $2.6\%$ & $0.4\%$~\cite{Chadwick2017} 	& $1.8\%$ & $0.5\%$ & $0.3\%$~\cite{Chadwick2017} \\ \hline
	$\delta\bar{\epsilon}_n^\mathrm{LAB}$ (MeV) 	& $2.5\%$ & $0.8\%$ & $0.5\%$~\cite{Capote2016}		& $0.7\%$ & $0.5\%$ & $1.7\%$~\cite{Neudecker2017} \\ \hline
	$\delta\overbar{M}_\gamma$ ($\gamma/f$) 	& $1.2\%$ & $0.5\%$ & $1.3\%$~\cite{Oberstedt2013} 	& $0.3\%$ & $0.3\%$ & $1.6\%$~\cite{Gatera2017} \\ \hline
	$\delta\bar{\epsilon}_\gamma$ (MeV) 		& $2.1\%$ & $0.9\%$ & $2.4\%$~\cite{Oberstedt2013} 	& $0.8\%$ & $0.7\%$ & $2.4\%$~\cite{Gatera2017} \\ \hline
     \end{tabular}
\caption{\label{tab:errors} Correlation between calculated $\bar{\nu}$ and the input average
  heavy fragment mass $\langle A_h\rangle$ for $\Unf$ and $\Punf$.
  Also listed are the biases for several prompt fission observables from using calculated mass yields
  of M{\"o}ller~\cite{Moller2015}, Sierk~\cite{Sierk2017}, as well as experimental uncertainties
  for reference.
}
\end{table*}
This correlation implies that, when all other input is kept constant, two mass yields with
heavy fragment peaks one mass unit apart will result in a $\bar{\nu}$ differing by about
$0.1\,n/f$. Very different peak widths $\sigma_{A_h}^2$ complicate the correlation.
We note that this analysis relies on the shape of the $\langle\TKE\rangle(A)$
we have chosen, but not on the overall $\langle\TKE\rangle$, which only shift the ordered pairs
and leave the correlation unaffected.

Also listed in Table~\ref{tab:errors} are the biases on the various prompt fission observables from the use
of calculated yields instead of experimental ones. We find that both the location of the
mass peak $\langle A_h\rangle$ and the width of the peak $\sigma^2_{A_h}$, where wider
peaks resulting in an increased yield near the $N=82$ shell closure, could result in a
slightly harder PFNS and PFGS. Specific discrete $\gamma$-ray intensities are also directly affected
by the choice of mass yields. The width of the mass peak was also
found to impact the correlation between the total kinetic energy of the
fragments and the average total prompt neutron multiplicity.

These correlations and derived biases will help inform future fission-yield models
and the de-excitation procedure. These calculations can be improved with self-consistent
$Y(A,Z)$ yields from Ref.~\cite{Moller2015} and $Y(A,\TKE)$ yields from Ref.~\cite{Sierk2017}.
In a future study, we plan to implement the exact fission-fragment mass yields into
the Hauser-Feshbach statistical-decay model and apply the effects of the experimental
mass and energy resolutions to the calculated results, instead of applying a mass
resolution to the input mass yields. Additional experimental data of the
fragment mass, charge, and kinetic energies at a variety of incident neutron energies,
such as Refs.~\cite{Duke2016,Meierbachtol2016}, would allow for a more critical
comparison of the calculated and experimental yields. Furthermore, measurements
of the fragment yields for exotic nuclei will improve our ability to benchmark
calculated yields outside the more well-studied actinide chains.
When calculating the prompt neutron and $\gamma$-ray emissions, several input parameters
are needed, but may not possess the proper energy dependence as there is no data available.
For example, the dependence of $\langle\TKE\rangle(A)$ on
incident neutron energy has only been determined for a limited number of
nuclei~\cite{Duke2016,Meierbachtol2016}. In addition, properties of the prompt $\gamma$ rays have
seldom been measured at higher incident neutron energies~\cite{Frehaut1983}, but
additional data may provide useful information about the spins of fission fragments at
these energies. Finally, measurements conducted by Naqvi {\it et al.}~\cite{Naqvi1986}
demonstrated that $\bar{\nu}(A)$ has a
distinct change in shape for higher incident neutron energies, but further experimental tests
of this would provide useful insight into the excitation energy sharing in fission.

Our results utilize theoretical methods to calculate fission observables
from scission to prompt neutron and $\gamma$-ray emissions, a step towards a predictive model of fission.
In general, we find that the use of calculated yields do not yet possess the precision needed
for very sensitive criticality estimates~\cite{Neudecker2016} or neutron correlation counting~\cite{Croft2015}.
However, it should be noted that the variance on $\bar{\nu}$ induced simply from
the differences in the experimental mass yields is already near the uncertainties of the
IAEA standards~\cite{Carlson2009}. For applications that do not require
this degree of accuracy, we find that the use of calculated mass yields and the prompt
particle emission through a Hauser-Feshbach treatment is invaluable, especially
where there is little to no experimental data as is the case in many nuclides participating
in the $r$-process~\cite{Cote2017}.
Furthermore, the prompt $\gamma$-ray observables appear less sensitive to the use of
calculated mass yields instead of experimental ones, suggesting that estimates of
$\gamma$-ray heating for reactor design could be done for nuclides without experimental
data using a combination of theoretical mass yields and a Hauser-Feshbach decay treatment,
as we have used here, and still satisfy the needed design uncertainties~\cite{Rimpault2012}.

\acknowledgements
The authors would like to thank A. G{\"o}{\"o}k for providing recent data
and T. Kawano, I. Stetcu, and M. White for useful conversations
on the subject. This work was supported by the Office of Defense Nuclear Nonproliferation
Research \& Development (DNN R\&D), National Nuclear Security Administration,
US Department of Energy. It was performed under the auspices of the National
Nuclear Security Administration of the US Department of Energy at Los Alamos
National Laboratory under Contract \protect{DEAC52-06NA25396}.




\end{document}